\documentclass[11pt]{article}

\usepackage{latexsym,amssymb,amsmath,amsthm,fullpage,times}
\def\Z{{\mathbb Z}}

\def\C{{\mathbb C}}

\newtheorem{theorem}{Theorem}
\newtheorem{lemma}{Lemma}
\newtheorem{proposition}{Proposition}

\newtheorem{definition}{Definition}

\begin{document}
\title{Quantum testers for hidden group properties\footnote{Research
partially supported by
the EU 5th framework programs QAIP IST-1999-11234 and
RAND-APX IST-1999-14036,
and by CNRS/STIC 01N80/0502 and 01N80/0607 grants,
and by OTKA T030059, T030132, and NWO-OTKA N34040 grants.\smallskip}}
\author{
Katalin Friedl\thanks{
CAI,
Hungarian Academy of Sciences,
H-1111 Budapest,
Hungary,
e-mail:  {\tt friedl@sztaki.hu}.
Work done while visiting CNRS--LRI, Orsay.
}
\and
Fr\'ed\'eric Magniez\thanks{
CNRS--LRI, UMR 8623 Universit\'e Paris--Sud, 
91405 Orsay, France,
e-mail:  \{{\tt magniez},{\tt santha}\}{\tt @lri.fr}.
}
\and
Miklos Santha${}^{\ddag}$
\and
Pranab Sen\thanks{
LRI, UMR 8623  Universit\'e Paris--Sud,
91405 Orsay, France,
e-mail:  {\tt pranab@lri.fr}.
}
}

\date{}

\newcommand{\ket}[1]{{|{#1}\rangle}}
\newcommand{\qft}{\mathrm{QFT}}
\newcommand{\size}[1]{\lvert #1 \rvert}
\newcommand{\norm}[1]{\left\lVert #1 \right\rVert}
\newcommand{\abs}[1]{\lvert #1 \rvert}
\newcommand{\dist}{\mathsf{dist}}
\newcommand{\prob}{\mathop{\mathsf{Pr}}}
\newcommand{\maj}{\mathop{\mathsf{Maj}}}
\renewcommand{\max}{\mathop{\mathsf{Max}}}
\renewcommand{\min}{\mathop{\mathsf{Min}}}
\newcommand{\group}[1]{{<}#1{>}}
\newcommand{\vect}[1]{\mathrm{Span}(#1)}
\renewcommand{\qedsymbol}{$\blacksquare$}
\newcommand{\lint}[1]{\lfloor #1 \rfloor}

\newcommand{\per}{\mathrm{Per}}
\newcommand{\equalcoset}{\mathrm{Range}}
\newcommand{\range}{\mbox{\rm CCR}} 
\newcommand{\longrange}{\mbox{\rm COMMON-COSET-RANGE}}
\newcommand{\totvar}[1]{\left\lVert #1 \right\rVert_1}

\newcommand{\kperiod}{\mbox{\rm LARGER-PERIOD($K$)}}

\newcommand{\encadre}[1]{
\centerline{\fbox{
\begin{minipage}{0.97\textwidth}
#1
\end{minipage}
}}}
\newsavebox{\fmbox}
\newenvironment{fmpage}[1]
      {\begin{lrbox}{\fmbox}\begin{minipage}{#1}}
      {\end{minipage}\end{lrbox}\fbox{\usebox{\fmbox}}}

\maketitle

\vspace*{-.7cm}

\begin{abstract}
We construct efficient or query efficient quantum property testers
for two existential group properties which have exponential
query complexity both for their   decision problem in the quantum
 and for their testing problem in the classical model of computing.
These are periodicity in groups and the common coset range property 
of  two functions having identical ranges within each coset of 
some normal subgroup.
Our periodicity tester is efficient in Abelian groups and generalizes,
in several aspects, previous periodicity testers. 
This is achieved by introducing a technique refining the
majority correction process widely used for proving
robustness of algebraic properties.
The periodicity tester in non-Abelian groups and the
common coset range tester are query efficient.
\end{abstract}

\section{Introduction}
In the paradigm of property testing one would like to decide
   whether an object has a global property by performing random
   local checks. The goal is
to distinguish with sufficient confidence
the objects which satisfy the property from those objects
that are far from having the property.
In this sense, property testing is a notion of approximation for
the corresponding decision problem.
%
Property testers, with a slightly different objective,
were first considered for programs under the name of self-testers.
Following the pioneering approach of Blum, Kannan, Luby and
Rubinfeld~\cite{bk95,blr93},
self-testers were constructed for programs
purportedly computing functions with some algebraic properties
such as linear functions, polynomial functions, and functions
satisfying some functional equations~\cite{blr93,rs96,rub99}.
The notion in its full generality was defined by Goldreich, Goldwasser
and Ron and successfully applied among others to graph
properties~\cite{ggr98,gr97}.
For surveys on property testing
see~\cite{gol98,ron00,kms00,fis01}.

Quantum computing is an extremely active research
area (for surveys see e.g.~\cite{rp00,aha98,pre98,nc00}),
where a growing trend is to cast
quantum algorithms in a group theoretical setting.
In this setting,
we are given a finite group $G$ and, besides the group operations,  
we also have at our disposal 
a function $f$ mapping  $G$ into a finite set.
The function $f$  can be queried via an oracle.
The complexity of an algorithm  is measured by the
number of queries ({\it i.e.} evaluations of the function $f$), and
also by the overall running time counting one query as one
computational step.
We say that an algorithm is
{\em query efficient} (resp. {\em efficient})
if  its query complexity (resp.  overall time complexity)
is polynomial in the logarithm of the order of $G$.
The most important unifying problem of group theory for the purpose
of quantum algorithms has turned out to be the
\textsc{Hidden Subgroup Problem (HSP)},
which can be cast in the following broad terms:
Let $H$ be a subgroup of $G$ 
such that $f$ is constant on each left coset of $H$
and distinct on different left cosets.
We say that $f$ {\em hides} the subgroup $H$.
The task is to determine the {\em hidden subgroup} $H$.

While no classical algorithm can solve this problem with polynomial
query complexity, the biggest success of
quantum computing until now is that it can be solved by a quantum
algorithm {\em efficiently} whenever $G$ is Abelian.
We will refer to this algorithm as the {\em standard algorithm}
for the HSP.
The main tool for this
solution is {\em Fourier sampling} based on
the (approximate) quantum Fourier transform for Abelian
groups which
can be efficiently implemented quantumly~\cite{kit95}.
Simon's xor-mask finding~\cite{sim97},
Shor's factorization
and discrete logarithm finding algorithms~\cite{sho97},
and Kitaev's algorithm~\cite{kit95} for the Abelian stabilizer problem
are all special cases of this general solution.
%
%
Fourier sampling was also successfully used to solve the closely related
\textsc{Hidden Translation Problem} (HTP). 
Here we are given two injective functions $f_{0}$ and $f_{1}$ from
an Abelian group $G$ to some finite set such that,
for some group element $u$, the equality
$f_{0}(x+u)=f_{1}(x)$ holds for every $x$. The task is to 
find the {\em translation} $u$.
Indeed, the HTP is an instance of the HSP in 
the semi-direct product $G\rtimes \Z_2$ where the
hiding function is $f(x,b)=f_{b}(x)$.
In that group $f$ hides the subgroup $H = \{(0,0), (u,1)\}$.
Ettinger and H{\o}yer~\cite{eh00} have shown that the HTP
can be solved in cyclic groups $G=\Z_n$ by a two-step procedure:
an efficient quantum algorithm followed by an
exponential classical stage without further queries.
They achieved this by applying Fourier sampling in the
Abelian direct product group $G\times\Z_2$.
In a recent work, we have shown~\cite{fmss02b} 
that HTP can be solved by an efficient quantum algorithm in 
some groups of fixed exponent, for instance
when $G=\Z_p^n$ for any fixed prime number $p$.
This gives a quantum polynomial
time algorithm for the HSP in $G\rtimes \Z_2$
using the quantum reduction of HSP to HTP~\cite{eh00}.
In strong opposition to these positive results,
a natural generalization of the HSP
has exponential quantum query complexity even in Abelian groups.
In this generalization, the function $f$ may
not be distinct on different cosets.
Indeed, the unordered database search problem can be reduced to
the decision problem whether a function on a cyclic group
has a non-trivial period or not.

Two different extensions of property testing were studied
recently in the quantum context.
The first approach consists in testing quantum devices by
classical procedures.
Mayers and Yao~\cite{MY98} have designed tests for
deciding if a photon source is perfect. These tests guarantee that
if a source passes them, it is adequate for the security of the
Bennett-Brassard~\cite{BB84} quantum key distribution protocol.
Dam, Magniez, Mosca and Santha~\cite{dmms00}
considered the design of testers for quantum gates.
They showed
the possibility of classically testing quantum processes
and they provided the first family of
classical tests allowing one
to estimate the reliability of quantum gates.

The second approach considers testing deterministic functions
by a quantum procedure.
Quantum testing of function families was
introduced by Buhrman, Fortnow, Newman, and R\"ohrig~\cite{bfnr02},
and they have constructed efficient quantum testers for several
properties.
One of their nicest contributions is that they
have considered the possibility that quantum testing
of periodicity might be easier than the corresponding
decision problem.
Indeed, they succeeded in giving a polynomial time quantum tester
for periodic functions over $\Z_{2}^{n}$.
They have also proved that any classical tester requires exponential
time for this task.
Independently and earlier, while working on the extension of the HSP
to periodic functions over $\Z$ which may be many-to-one in
each period, Hales and Hallgren~\cite{hh00,halesphd}
have given the essential ingredients for constructing a polynomial
time quantum tester for periodic functions over the cyclic group
$\Z_{n}$.
But contrarily to~\cite{bfnr02}, their result is not stated
in the testing context.

In this work, we construct efficient or query efficient
quantum testers for two {\em hidden group properties},
that is, existential
properties over groups whose decision problems have
exponential quantum query complexity.
We also introduce a new technique in the analysis of quantum
testers.

Our main contribution is a generalization
of the periodicity property studied in \cite{hh00,bfnr02}.
For any finite group $G$ and any normal subgroup $K$,
a function $f$ satisfies the property $\kperiod$ if
there exists a normal subgroup $H>K$ for which
$f$ is $H$-periodic ({\em i.e.} $f(xh) = f(x)$ for all
$x \in G$ and $h \in H$).
For this property, we give an efficient tester
whenever $G$ is Abelian (\textbf{Theorem~\ref{theorem:ab-period}}).
This result generalizes the previous periodicity testers
in three aspects.
First, we work in any finite Abelian group $G$, while
previously only $G = \Z_n$~\cite{hh00} and $G = \Z_2^n$~\cite{bfnr02}
were considered.
Second, the property we test is parametrized by
some known normal subgroup $K$, while previously only the
case $K=\{0\}$ was considered.
Third, our query complexity is only linear in the inverse of the
distance parameter, whereas the previous
works have a quadratic dependence.
These improvements are possible due to our more transparent analysis.
We refine the standard method of classical testing, which
consists in showing that a function $f$ that passes
the test can be corrected into another function $g$
that has the desired property,
and which is close to $f$.
The novelty of our approach is that here the correction is not done
directly; it involves an intermediate correction via a probabilistic
function.

The main technical ingredient of the periodicity test in
Abelian groups is efficient {\em Fourier sampling}.
This procedure remains  a powerful tool also in non-Abelian groups.
Unfortunately, currently no  efficient implementation 
is known for it in general groups.
Therefore, when dealing with non-Abelian groups, our aim is to
construct query efficient testers.
We construct query efficient testers for two properties.
First, we show that the  tester used for $\kperiod$  in Abelian groups
yields a query efficient tester when $G$ is any finite group and
$K$ any normal subgroup (\textbf{Theorem~\ref{theorem:period}}).
Second, we study in any finite group $G$ the property $\longrange(k,t)$
(for short $\range(k,t)$) related to the HTP.
Let $f, g$ be two functions from $G$ to a finite
set $S$. By definition, $(f, g)$ satisfies $\range(k,t)$
if $f$ and $g$ have identical ranges within each coset for a
normal subgroup $H \unlhd G$ of size at most $k$, and which is the normal
closure of a subgroup generated by at most $t$ elements.
The heart of the tester for $\range(k,t)$ is again Fourier sampling
applied in the direct product group $G\times \Z_2$.
Our tester is query efficient in any group
if $k$ is polylogarithmic in the size of the group
(\textbf{Theorem~\ref{theorem:cosetrange}}). 

Different lower bounds can be proven on the query 
complexity of $\range(k,t)$.
One observes easily that unordered database search can be
reduced to $\range(k,1)$ in Abelian groups of exponent $k$, and therefore
$\range(k,1)$ is quantumly exponentially hard to decide.
Moreover, we show that classical testers also require
an exponential number of queries for this problem even if $k$ is constant
(\textbf{Theorem~\ref{thm:lowerbound}}).
We show this by adapting the techniques of~\cite{bfnr02},
who proved the analogous result for classical testers for periodicity.

\section{Preliminaries}
\subsection{Fourier sampling over Abelian groups}
For a finite set $D$, let the {\em uniform superposition
over $D$} be
$\ket{D}=\tfrac{1}{\sqrt{\size{D}}}\sum_{x\in D}\ket{x}$,
and for a function $f$ from $D$ to a finite set $S$,
let the {\em uniform superposition of $f$} be
$\ket{f}=\tfrac{1}{\sqrt{\size{D}}}\sum_{x\in D}\ket{x}\ket{f(x)}$.
For two functions $f,g$ from $D$ to $S$,
their {\em distance} is
$\dist(f,g)=\size{\{x\in D : f(x)\neq g(x)\}}/\size{D}$.
The following proposition describes the relation between the
distance of two functions and the distance between their
uniform superpositions. In this paper, $\norm{\cdot}$ denotes the
$L_2$-norm of a vector.
\begin{proposition}\label{prop:distance}
For functions $f,g$ defined on the same finite set,
$\dist(f,g)=\frac{1}{2}\norm{\ket{f}-\ket{g}}^2$.
\end{proposition}

Let $G$ be a finite Abelian group and $H\leq G$ a subgroup.
The coset of $x \in G$ with respect to $H$ is denoted by
$x+H$. We use the notation $\group{X}$ for
the subgroup generated by a subset $X$ of $G$.
We identify  with $G$ the set $\widehat{G}$ of characters
of $G$, via some fixed isomorphism $y\mapsto \chi_{y}$.
The {\em orthogonal of $H\leq G$} is
defined as $H^{\perp}=\{y\in G : \forall h\in H, \chi_{y}(h)=1\}$,
and we set
$\ket{H^\perp(x)}=\sqrt{\tfrac{\size{H}}{\size{G}}}
\sum_{y\in H^\perp}\chi_{y}(x)\ket{y}.$
The {\em quantum Fourier transform} over $G$, $\qft_G$, is
the unitary transformation defined as follows: For every $x\in G$, 
$\qft_G \ket{x} =
\tfrac{1}{\sqrt{\size{G}}}\sum_{y \in G} \chi_{y}(x)\ket{y}$.
The main property about $\qft_G$ that we use is that it
maps the uniform superposition on the coset $x+H$ to
the uniform superposition on $H^\perp$, with appropriate phases.
\begin{proposition}\label{prop:abelian-fourier}
Let $G$ be a finite Abelian group,
$x\in G$ and $H\leq G$.
Then $\ket{x+H}\xrightarrow{\qft_G}\ket{H^\perp(x)}$.
\end{proposition}

The following well known quantum Fourier sampling
algorithm will be used as a building block in our quantum testers.
In the algorithm, $f:G\rightarrow S$ is given by a quantum oracle.
\begin{center}
\begin{fmpage}{10.5cm}
\textbf{Fourier sampling}${}^{f}(G)$
\begin{enumerate}
\setlength{\itemsep}{-2mm}
\item Create zero-state $\ket{0}_{G}\ket{0}_{S}$.
\item Create the superposition 
      $\tfrac{1}{\sqrt{\size{G}}}\sum_{x\in G}\ket{x}$
      in the first register.
\item Query function $f$.
\item Apply $\qft_{G}$ on the first register.
\item Observe and then output the first register.
\end{enumerate}
\end{fmpage}
\end{center}
The above algorithm is actually the main ingredient
for solving the HSP on Abelian groups
with hiding function $f$.

\subsection{Property testing}
Let $D$ and $S$ be two finite sets
and let~$\mathcal{C}$ be a family of functions from $D$
to $S$. Let
$\mathcal{F}\subseteq\mathcal{C}$ be the subfamily of
functions of interest, that is, the set of functions
possessing the desired property.
In the testing problem, one is interested in distinguishing
functions $f:D\to S$, given by an oracle, which belong to
$\mathcal{F}$, from functions which are far from every function
in $\mathcal{F}$.
\begin{definition}[$\delta$-tester]
Let $\mathcal{F}\subseteq\mathcal{C}$ and $0\leq\delta<1$.
A \emph{quantum} (resp. \emph{probabilistic}) \emph{$\delta$-tester}
for $\mathcal{F}$ on $\mathcal{C}$
is a quantum (resp. probabilistic) oracle Turing machine $T$
such that, for every $f\in\mathcal{C}$,\\
\indent 1. if $f\in\mathcal{F}$
then $\prob[T^{f}\ \text{accepts}]=1$,\\
\indent 2.
if $\dist(f,\mathcal{F})>\delta$
then $\prob[T^{f}\ \text{rejects}] \geq 2/3$,\\
where the probabilities are taken over the observation results
(resp. the coin tosses) of $T$.
\end{definition}
By our definition, a tester always accepts functions
having the property $\mathcal{F}$.
We may also consider testers with {\em two-sided error},
where this condition is relaxed, and one requires only that the tester
accept functions from $\mathcal{F}$ with probability at least 2/3.
Of course, the choice of the success probability $2/3$
is arbitrary, and can be replaced by $\gamma$, for any constant
$1/2 < \gamma < 1$.

\section{Periodicity}
In this section, we design quantum testers for testing periodicity 
of functions from
a finite group $G$ to a finite set $S$.
For a normal subgroup $H\unlhd G$, a function $f:G\rightarrow S$
is {\em $H$-periodic} if for all $x\in G$ and $h\in H$,
$f(xh)=f(x)$.
Notice that our definition describes formally right
$H$-periodicity, but this coincides with left $H$-periodicity
since $H$ is normal.
The set of $H$-periodic functions is denoted by $\per(H)$.
For a known normal subgroup $H$, testing if $f\in\per(H)$ can 
be easily done classically by sampling
random elements $x\in G$ and $h\in H$, and verifying that
$f(xh)=f(x)$.
On the other hand, testing if a function has a non-trivial period
is classically hard even in $\Z_{2}^{n}$~\cite{bfnr02}.
The main result of this section is that we can test query efficiently
by a quantum algorithm an even more general property:
Does a function have a strictly larger period than a
known normal subgroup
$K\unlhd G$? Indeed, we test the family
$$\kperiod=\{f:G\rightarrow S\ \mid\ \exists
H\unlhd G,\ H>K \text{ and $f$ is $H$-periodic}\}.$$
Moreover when $G$ is Abelian, our tester is efficient.


For the sake of clarity we first present the result
for Abelian groups.
This enables us to highlight the new technique that we use.
The standard way to ensure that the functions the tester accepts
with high probability are close to functions having
the desired property,
is based on a direct correction process.
This process has to produce a corrected function
which has the desired property, and is close to the original function.
This is the approach taken by~\cite{hh00,bfnr02}.
The novelty of our approach is that the correction is not done
directly; it involves an intermediate corrected probabilistic
function.
This two-step process makes a more refined and cleaner analysis
possible, and allows us to prove that our tester works
in any finite group, whereas  previous works only considered the groups
$\Z_n$~\cite{hh00} and $\Z_2^n$~\cite{bfnr02}. Moreover, the
query complexity of our algorithm turns out to be
linear in the inverse of the distance parameter, unlike
the quadratic dependence of the previous works.

\subsection{Abelian case}\label{sec:abelian}
In this subsection, we give our algorithm for testing periodicity
in abelian groups.
Theorem~\ref{theorem:ab-period} below states that this algorithm
is efficient. 
The algorithm assumes that $G$ has
an efficient exact quantum Fourier transform.
When $G$ only has an efficient approximate quantum
Fourier transform, the algorithm has two-sided error.
Efficient implementations of approximate quantum Fourier transforms
exist in every finite Abelian group~\cite{kit95}.

\begin{center}
\begin{fmpage}{9cm}
\textbf{Test Larger period${}^{f}(G,K,\delta)$}
\begin{enumerate}
\setlength{\itemsep}{-2mm}
\item $N\leftarrow 4\log(\size{G})/\delta$.
\item For $i=1,\ldots,N$ do
$y_{i}\leftarrow\mbox{\textbf{Fourier sampling}}^{f}(G)$.
\item Accept iff $\ \group{y_{i}}_{1\leq i\leq N} < K^\perp$.
\end{enumerate}
\end{fmpage}
\end{center}

\begin{theorem}\label{theorem:ab-period}
For a finite set $S$, finite Abelian group $G$,
subgroup $K \le G$, and $0<\delta<1$,
{\rm\textbf{Test Larger period}}$(G,K,\delta)$ is
a $\delta$-tester for $\kperiod$ on the family of all functions
from $G$ to $S$,
with $O(\log(\size{G})/\delta)$ query complexity
and $(\log(\size{G})/\delta)^{O(1)}$ time complexity.
\end{theorem}


Let $S$ be a finite set and $G$ a finite Abelian group.
We describe now the ingredients of our two-step correction process.
First, we generalize the notion of uniform superposition of a function
to uniform superposition of a probabilistic function.
By definition,
a {\em probabilistic function} is a mapping $\mu: x\mapsto \mu_{x}$
from the domain $G$ to probability distributions on $S$.
For every $x \in G$, define the unit $L_{1}$-norm vector
$\ket{\mu_{x}}=\sum_{s\in S}\mu_{x}(s)\ket{s}$.
Then the uniform superposition of $\mu$ is defined as
$\ket{\mu}=\tfrac{1}{\sqrt{\size{G}}}\sum_{x\in G}\ket{x}\ket{\mu_{x}}$.
Notice that $\ket{\mu}$ has unit $L_{2}$-norm when $\mu$ is a
(deterministic) function,
otherwise its $L_{2}$-norm is smaller.

A function $f:G\rightarrow S$ and a subgroup $H\leq G$ naturally
define an $H$-periodic probabilistic function $\mu^{f,H}$,
where $\mu_{x}^{f,H}(s)=\frac{\size{f^{-1}(s)\cap (x+H)}}{\size{H}}$.
The value $\mu_{x}^{f,H}(s)$ is the proportion of elements in the
coset $x+H$ where $f$ takes the value $s$.
Observe that when $f$ is $H$-periodic $\ket{\mu^{f,H}}=\ket{f}$,
and so $\norm{\ket{\mu^{f,H}}}=1$, otherwise
$\norm{\ket{\mu^{f,H}}}<1$.

First, we give the connection between
the probability that
$\mbox{\textbf{Fourier sampling}}$ outputs an element outside
$H^{\perp}$, and
the distance between $\ket{f}$ and $\ket{\mu^{f,H}}$.
\begin{lemma}\label{lemma:period1}
$\norm{\ket{f}-\ket{\mu^{f,H}}}^2=
\prob[\mbox{\rm\textbf{Fourier sampling}}^f(G)\text{ outputs }
y\not\in H^\perp].
$
\end{lemma}
\begin{proof}
The probability term is
$\norm{\frac{1}{\sqrt{\size{G}}}\sum_{x\in G}
\ket{\{0\}^{\perp}(x)}\ket{f(x)}
-\frac{1}{\sqrt{\size{G}\size{H}}}\sum_{x\in G}
\ket{H^{\perp}(x)}\ket{f(x)}}^2$,
since $y\not\in H^{\perp}$ iff $y\in\{0\}^{\perp}-H^{\perp}$.
We apply the inverse quantum Fourier transform $\qft_G^{-1}$, which is
$L_2$-norm preserving, to the first register in the above expression.
The probability becomes
$\norm{\ket{f}-\frac{1}{\sqrt{\size{G}\size{H}}}\sum_{x\in G}
\ket{x+H}\ket{f(x)}}^2$,
using  Proposition~\ref{prop:abelian-fourier}.
Changing the variables, the second term inside the norm is
$$
\frac{1}{\sqrt{\size{G}}}\sum_{x\in G}\ket{x}\frac{1}{\size{H}}
\sum_{h\in H}\ket{f(x-h)}
=\frac{1}{\sqrt{\size{G}}}\sum_{x\in G}\ket{x}\frac{1}{\size{H}}
\sum_{h\in H}\ket{f(x+h)},
$$
where the equality holds because $H$ is a subgroup of $G$.
We conclude by observing that, by definition of $\mu^{f,H}$,
$\frac{1}{\size{H}}\sum_{h\in H}\ket{f(x+h)}
=\sum_{s\in S}\mu_{x}^{f,H}(s)\ket{s} = \ket{\mu_x^{f,H}}$.
\end{proof}

Second, we give the connection between $\dist(f,\per(H))$
and the distance between $\ket{f}$ and $\ket{\mu^{f,H}}$.
\begin{lemma}\label{lemma:period2}
$\dist(f,\per(H))\leq 2\norm{\ket{f}-\ket{\mu^{f,H}}}^2.$
\end{lemma}
\begin{proof}
It will be useful to rewrite $\ket{f}$ as
a probabilistic function
$\frac{1}{\sqrt{\size{G}}}\sum_{x\in G}\ket{x}\sum_{s\in S}
\delta_{x}^f(s)\ket{s}$, where
$\delta_{x}^f(s)=1$ if $f(x)=s$ and $0$ otherwise.
Let us define the $H$-periodic function $g:G\rightarrow S$
by $g(x)=\maj_{h\in H}{f(x+h)}$, where ties are decided arbitrarily.
In fact, $g$ is the correction of $f$ with respect
to $H$-periodicity.
Proposition~\ref{prop:distance} and the $H$-periodicity
of $g$ imply
$\dist(f,\per(H))\leq\frac{1}{2}\norm{\ket{f}-\ket{g}}^2$.
We will show that
$\norm{\ket{g}-\ket{\mu^{f,H}}}\leq\norm{\ket{f}-\ket{\mu^{f,H}}}$.
This will allow us to prove the desired statement
using the triangle inequality.
%
Observe that for any function $h:G\rightarrow S$, we have
\begin{equation}
\label{eq:deltamu}
\norm{\ket{h}-\ket{\mu^{f,H}}}^2=
\frac{1}{\size{G}}\sum_{x\in G}
\sum_{s\in S}\abs{\delta_{x}^h(s)-\mu^{f,H}_{x}(s)}^2.
\end{equation}
Moreover for every $x\in G$, one can establish
\begin{equation}
\label{eq:majorisation}
\begin{array}{c c l}
\sum_{s\in S}\abs{\delta_{x}^g(s)-\mu_{x}^{f,H}(s)}^2
&=&\abs{1-\mu_{x}^{f,H}(g(x))}^2+\sum_{s\neq g(x)}
    (\mu_{x}^{f,H}(s))^2\\
&=&1+\sum_{s\in S} (\mu_{x}^{f,H}(s))^2 - 2\mu_{x}^{f,H}(g(x))\\
&\leq&1+\sum_{s\in S} (\mu_{x}^{f,H}(s))^2 - 2\mu_{x}^{f,H}(f(x))\\
&=&\sum_{s\in S}\abs{\delta_{x}^f(s)-\mu_{x}^{f,H}(s)}^2,
\end{array}
\end{equation}
where the inequality follows from
$\mu_{x}^{f,H}(f(x))\leq \mu_{x}^{f,H}(g(x))$,
which in turn follows immediately from the definition of $g$.

{From} (\ref{eq:deltamu}) and (\ref{eq:majorisation}) we get that
$\norm{\ket{g}-\ket{\mu^{f,H}}}\leq\norm{\ket{f}-\ket{\mu^{f,H}}}$,
which completes the proof.
\end{proof}

Lemmas~\ref{lemma:period1} and \ref{lemma:period2}
together can be interpreted
as the robustness~\cite{rs96,rub99}
in the quantum context~\cite{dmms00} of the property that
\textbf{Fourier sampling}${}^f(G)$ outputs only $y\in H^\perp$:
if $f$ does not satisfy exactly the property but
with error probability less than $\delta$, then $f$ is $2\delta$-close
to a function that satisfies exactly the property.
Using that fact, we can now prove Theorem~\ref{theorem:ab-period}.
\begin{proof}[Proof of Theorem~\ref{theorem:ab-period}]
If $f\in\kperiod$, that is
$f$ is $H$-periodic for some $H>K$,
then the quantum state before the observation of
$\mbox{\textbf{Fourier sampling}}^f(G)$ is
\begin{eqnarray*}
(\qft_G \otimes I) 
     \left(\frac{1}{\sqrt{\size{G}}} \sum_{x\in G} \ket{x} 
     \ket{f(x)}\right) 
&=&
(\qft_G \otimes I) 
     \left(\frac{1}{\sqrt{\size{G}\size{H}}} \sum_{x\in G} \ket{x+H} 
     \ket{f(x)}\right) \\
&=&
\frac{1}{\sqrt{\size{G}\size{H}}}\sum_{x\in G} \ket{H^{\perp}(x)}
\ket{f(x)}.
\end{eqnarray*}
Above, $I$ denotes the $\size{S} \times \size{S}$
identity matrix.
Therefore, $\mbox{\textbf{Fourier sampling}}^f(G)$
only outputs elements in $H^\perp$.
Since $H^{\perp}<K^{\perp}$, the test always accepts.

Let $f$ be now $\delta$-far from $\kperiod$.
Then for every $H>K$ $\dist(f,\per(H))>\delta$, and by
Lemmas~\ref{lemma:period1} and~\ref{lemma:period2},
$\prob[\mbox{\textbf{Fourier sampling}}^f(G)\text{ outputs }
y\not\in H^\perp]>\delta/2$.
Using these inequalities, we can upper bound the
acceptance probability of the test as follows.
\begin{eqnarray*}
\prob[\group{y_{i}}_{1\leq i\leq N} < K^{\perp}] 
&=&
\prob[\exists H>K, \group{y_{i}}_{1\leq i\leq N}\leq H^{\perp}]\\
&=&
\prob[\exists x\in G-K, {y_{i}}\in
\group{K,x}^{\perp}\  \text{ for } {1\leq i\leq N}]\\
&\leq& 
\size{G} \cdot \left(\max_{H>K}\left\{
\prob[\mbox{\textbf{Fourier sampling}}^f(G)\text{ outputs }
y\in H^\perp]\right\}\right)^{N}\\
&<& \size{G}(1-\delta/2)^{N} \leq 1/3.
\end{eqnarray*}
\end{proof}

\subsection{General case}
\label{subsec:nonabelian}

We start with a few definitions.
For any $d \times d$ matrix $M$, define
$\ket{M} = \sqrt{d} \sum_{1\leq i,j\leq d} M_{i,j} \ket{M,i,j}$.
Let $G$ be any finite group and
let $\widehat{G}$ be a complete set of finite dimensional
inequivalent irreducible unitary representations of $G$.
Thus, for any $\rho\in\widehat{G}$ of dimension $d_\rho$
and $x\in G$, $\ket{\rho(x)}=\sqrt{d_\rho}
\sum_{1\leq i,j\leq d_\rho}(\rho(x))_{i,j}\ket{\rho,i,j}$.
The {\em quantum Fourier transform}
over $G$ is the unitary transformation
defined as follows: For every $x\in G$, 
$\qft_G \ket{x} =
\tfrac{1}{\sqrt{\size{G}}}\sum_{\rho \in \widehat{G}} \ket{\rho(x)}$.
For any $H\unlhd G$ set
$H^{\perp}=\{\rho\in\widehat{G} : \forall h\in H,
\rho(h)=I_{d_{\rho}}\}$,
where $I_{d_{\rho}}$ is the $d_{\rho} \times d_{\rho}$ identity matrix.
Let $\ket{H^\perp(x)}=\sqrt{\frac{\size{H}}{\size{G}}}
\sum_{\rho\in H^\perp}\ket{\rho(x)}$. 

\begin{proposition}\label{prop:normal-fourier}
Let $G$ be a finite group,
$x\in G$ and $H\unlhd G$.  Then
$\ket{xH}\xrightarrow{\qft_G}
\ket{H^\perp(x)}$.
\end{proposition}
\begin{proof}
We first prove that when $H$ is normal, the matrix
$L=\sum_{h\in H}\rho(h)$ is $\size{H}\cdot
I_{d_\rho}$ if $\rho\in H^\perp$, and $0$ otherwise.
By definition of $H^\perp$, the condition
$\rho\in H^\perp$ implies $\rho(h)=I_{d_\rho}$ for every $h\in H$,
which gives the first part of the above.
Now suppose that $\rho\not\in H^\perp$.
Observe that since $H$ is normal, $L$ commutes with $\rho(x)$
for every $x\in G$. Therefore according to
Schur's lemma (see for instance~\cite[Chap.~2, Prop.~4]{ser77}),
$L=\lambda\cdot I_{d_\rho}$ for some $\lambda\in\C$.
Since $\rho\not\in H^\perp$,
we can pick some $h\in H$ such that $\rho(h)\neq I_{d_\rho}$;
then applying $\rho(h)\cdot L=L$ gives a contradiction if
$\lambda\neq 0$. This proves the second part of the above.

We now complete the proof of the proposition as follows.
\begin{eqnarray*}
\qft_G \ket{xH} &=& \frac{1}{\sqrt{\size{H}}}
                     \sum_{h \in H} \qft_G \ket{xh} \\
                 &=& \frac{1}{\sqrt{\size{H} \size{G}}}
                     \sum_{h \in H} \sum_{\rho \in \widehat{G}}
                     \ket{\rho(xh)} \\
                 &=& \frac{1}{\sqrt{\size{H} \size{G}}}
                     \sum_{\rho \in \widehat{G}}
                     \ket{\sum_{h \in H} \rho(xh)} \\
                 &=& \frac{1}{\sqrt{\size{H} \size{G}}}
                     \sum_{\rho \in \widehat{G}}
                     \ket{\rho(x) \cdot L} \\
                 &=& \sqrt{\frac{\size{H}}{\size{G}}}
                     \sum_{\rho \in H^\perp} \ket{\rho(x)} \\
                 &=& \ket{H^\perp(x)},
\end{eqnarray*}
where the penultimate equality follows from the above property
of the matrix $L$.
\end{proof}

We now give our algorithm for testing periodicity
in a general finite group $G$. In the algorithm,
$\textbf{Fourier sampling}^f(G)$ is as before, except that we
only observe the representation $\rho$, and not the indices
$i,j$. Thus, the output of $\textbf{Fourier sampling}^f(G)$ is
an element of $\widehat{G}$. $K$ is assumed to be a normal
subgroup of $G$. For any $\rho \in \widehat{G}$, $\ker{\rho}$ denotes its 
kernel.

\begin{center}
\begin{fmpage}{9cm}
\textbf{Test Larger period${}^{f}(G,K,\delta)$}
\begin{enumerate}
\setlength{\itemsep}{-2mm}
\item $N\leftarrow 4\log(\size{G})/\delta$.
\item For $i=1,\ldots,N$ do
$\rho_{i}\leftarrow\mbox{\textbf{Fourier sampling}}^{f}(G)$.
\item Accept iff $\cap_{1\leq i\leq N} \ker{\rho_i} > K$.
\end{enumerate}
\end{fmpage}
\end{center}

We now prove the robustness of the property that
$\textbf{Fourier sampling}^f(G)$ outputs only $\rho \in H^\perp$,
for any finite group $G$, normal subgroup $H$ and
$H$-periodic function $f$.
This robustness corresponds to Lemmas~\ref{lemma:period1} and
\ref{lemma:period2} of the Abelian case.
\begin{lemma}\label{lemma:normal}
Let $f:G\rightarrow S$ and $H\unlhd G$. Then
$$\dist(f,\per(H)) \leq
2\cdot\prob[\mbox{\rm\textbf{Fourier sampling}}^f(G)\text{ outputs }
\rho\not\in H^\perp].$$
\end{lemma}
\begin{proof}
The proof has the structure of the Abelian case
(see Lemmas~\ref{lemma:period1} and \ref{lemma:period2}).
Define $\ket{\mu^{f,H}}$ in the same way.
Observe that Lemma~\ref{lemma:period2} is true in a general finite 
group. The proof of Lemma~\ref{lemma:period1}
follows the one for the Abelian case. The only difference is that
we have to use Proposition~\ref{prop:normal-fourier} instead of
Proposition~\ref{prop:abelian-fourier}.
\end{proof}

Our second theorem states that {\rm\textbf{Test Larger period}}
is a query efficient tester for {\kperiod} for any finite
group $G$.
\begin{theorem}\label{theorem:period}
For a finite set $S$, finite group $G$,
normal subgroup $K\unlhd G$, and $0<\delta<1$,
{\rm\textbf{Test Larger period}}$(G,K,\delta)$ is
a $\delta$-tester for $\kperiod$ on the family of all functions
from $G$ to $S$,
with $O(\log(\size{G})/\delta)$ query complexity.
\end{theorem}
\begin{proof}
The proof is similar to that of the Abelian case. Note that,
while upper bounding the acceptance probability of the test
when $f$ is $\delta$-far from $\kperiod$, one has to consider
only those normal subgroups $H$ of the form
$H = \mbox{{\rm Normal-closure}}(\group{K,x})$, where $x$ ranges
over $G-K$.
\end{proof}

\section{Common Coset Range}
In this section, $G$ denotes a finite group and $S$ a finite set.
Let $f_0, f_1$ be functions from $G$ to $S$.
For a normal subgroup $H\unlhd G$, we say that  $f_{0}$ and $f_{1}$ are
{\em $H$-similar} if on all cosets of $H$
the ranges of $f_{0}$ and $f_{1}$ are the same,
that is,  the multiset equality
$f_{0}(xH)=f_{1}(xH)$ holds for every $x\in G$.
The couple of functions $(f_0, f_1)$ 
can equivalently be considered as a
single function $f : G \times \Z_2\rightarrow S$,
where by definition $f(x,b) = f_b(x)$.
We will use $f$ for $(f_{0},f_{1})$ when it is convenient in the
coming discussion.
We denote by $\equalcoset(H)$ the set of functions $f$ such that
$f_{0}$ and $f_{1}$ are $H$-similar.
We say that $H$ is {\em $t$-generated}, for some positive integer $t$,
if it is the normal closure of a subgroup generated by at most $t$
elements.
The aim of this section is to establish that for any 
positive integers $k$ and $t$, 
the family $\longrange(k,t)$ (for short $\range(k,t)$), defined as
$$\range(k,t)=\{f:G\times\Z_{2}\rightarrow S \ \mid\ \exists
H \unlhd G, \size{H} \le k, \text{$H$ is $t$-generated},
\text{$f_{0}$ and $f_{1}$ are $H$-similar}\},$$
can be tested by the following quantum test.
Note that a subgroup of size $k$ is always generated  by
at most $\log{k}$ elements, therefore we always assume that $t\leq\log k$.
\begin{center}
\begin{fmpage}{14cm}
\textbf{Test Common coset range${}^{f}(G,k,t,\delta)$}
\begin{enumerate}
\setlength{\itemsep}{-2mm}
\item $N\leftarrow 2kt\log(\size{G})/\delta$.
\item For $i=1,\ldots,N$ do $(\rho_{i},b_{i})\leftarrow
\mbox{\textbf{Fourier sampling}}^{f}(G\times\Z_{2})$.
\item Accept iff $\quad\exists H \unlhd G,\size{H} \le k,
\text{$H$ is $t$-generated}\quad 
\forall i\ (b_{i}=1\implies \rho_{i}\not\in H^{\perp})$.
\end{enumerate}
\end{fmpage}
\end{center}

We first prove the robustness of the property that
when $\textbf{Fourier sampling}^{f}(G\times\Z_{2})$ outputs $(\rho,1)$,
where $G$ is any finite group, $H \unlhd G$ and $f \in \equalcoset(H)$,
then $\rho$ is not in $H^{\perp}$.
\begin{lemma}\label{lemma:equalcoset}
Let $S$ be a finite set and $G$ a finite  group.
Let $f:G \times \Z_2 \rightarrow S$
and $H\unlhd G$. Then
$\dist(f,\equalcoset(H)) \leq
\size{H}\cdot
\prob[\mbox{\rm\textbf{Fourier sampling}}^{f}(G\times\Z_{2})
\text{ outputs } (\rho,1) \text{ such that } \rho \in H^\perp]$.
\end{lemma}
\begin{proof}
We use the notations of Section~\ref{sec:abelian}
for $f_0$ and $f_1$.
We define
$\ket{f,H} =
\frac{1}{\sqrt{2}}(\ket{\mu^{f_0,H}}-
\ket{\mu^{f_1,H}})$,
and the multiplicity functions
$m_{x}^{f_{b},H}=\size{H}\cdot \mu_{x}^{f_{b},H}$.

First, we prove that
$\dist(f,\equalcoset(H))\leq \norm{\ket{f,H}}^2\cdot |H|/2$.
For this, we define a function
$g_1:G\rightarrow S$, the correction of $f_1$.
The definition is done according to the cosets of $H$ in $G$.
For every $x\in G$ and $s \in S$, the function $g_1$ remains identical
to $f_1$ in $\min\{m^{f_0,H}_{x}(s),m^{f_1,H}_{x}(s)\}$
elements of $xH$, and the value of $g_1$ at those elements is $s$; 
at the remaining
elements of $xH$, the values of $g_1$ are defined 
so as to make the multisets $f_0(xH)$ and $g_1(xH)$ equal.
If we define $g_0 = f_0$ then
clearly $g = (g_0,g_1)$ is in $\equalcoset(H)$,
and $\dist(f,g) = \dist(f_1,g_1)/2$.
Since in every coset $xH$, $f_1$ and $g_1$ have
different values in
$\sum_{s \in S}\abs{m^{f_0,H}_{x}(s) -  m^{f_1,H}_{x}(s)}/2$
elements, we have
$\dist(f,g) = \frac{1}{4\size{G}}\sum_{x\in G/H}
\sum_{s \in S} \abs{m^{f_0,H}_{x}(s) -  m^{f_1,H}_{x}(s)}$.
The right hand side becomes $\norm{\ket{f,H}}^2 \cdot \size{H}/2$
if we replace the terms
$\abs{m^{f_0,H}_{x}(s)-m^{f_1,H}_{x}(s)}$ by
their respective squared values.
This can only increase it since the values $m^{f_b,H}_{x}(s)$
are integers. Thus, $\dist(f,g) \leq \norm{\ket{f,H}}^2\cdot |H|/2$. 

We now prove that
$\norm{\ket{f,H}}^2=2\cdot
\prob[\mbox{\textbf{Fourier sampling}}^f\text{ outputs } (\rho,1)
\text{ such that } \rho\in H^\perp]$.
The probability term is
$\norm{\frac{1}{2\sqrt{\size{H}\size{G}}}\sum_{x\in G}
\ket{H^{\perp}(x)} \ket{1}(\ket{f_0(x)}-\ket{f_1(x)})}^2$.
We apply the inverse quantum Fourier transform 
$\qft_G^{-1}$, which is $L_2$-norm preserving, to the first register
in the above expression.
Using Proposition~\ref{prop:normal-fourier} and the fact that
$H$ is a subgroup of $G$,
the probability becomes
$\norm{\frac{1}{2\size{H}\sqrt{\size{G}}}
\sum_{x\in G} \sum_{h\in H}
\ket{x}\ket{1}(\ket{f_0(xh)}-\ket{f_1(xh)})}^2$.
Now one can conclude the above statement and hence the lemma,
since by definition of $m^{f_{b},H}$,
the equality
$\sum_{h\in H}\ket{f_b(xh)}
=\sum_{s\in S} m_{x}^{f_b,H}(s)\ket{s}$ holds.
\end{proof}

Our next theorem implies that {$\range(k,t)$} is query efficiently
testable when $k$ is polynomial in $\log\size{G}$.
\begin{theorem}\label{theorem:cosetrange}
For any finite set $S$, finite group $G$, 
integers $k\geq 1$, $1\leq t\leq\log k$, and $0<\delta<1$,
{\rm\textbf{Test Common coset range}}$(G,k,t,\delta)$ is
a $\delta$-tester for $\range(k,t)$ on the family of all functions
from $G\times\Z_{2}$ to $S$,
with $O(k t\log(\size{G})/\delta)$ query complexity.
\end{theorem}
\begin{proof}
First consider the case $f\in\range(k,t)$, that is
$f$ is in $\equalcoset(H)$ for some
$H \unlhd G$, $|H| \le k$ and $H$ is $t$-generated.
{From} the proof of Lemma~\ref{lemma:equalcoset}, we see that whenever
$\mbox{\textbf{Fourier sampling}}^{f}(G\times\Z_{2})$
outputs an element $(\rho,1)$, then $\rho\not\in H^{\perp}$.
Thus the test always accepts.

Now, let $f:G\rightarrow S$ be $\delta$-far from $\range(k,t)$
and let $H$ be a $t$-generated normal subgroup of size at most $k$. Then
$\dist(f,\equalcoset(H))>\delta$ 
and by Lemma~\ref{lemma:equalcoset},
$\prob[\mbox{\textbf{Fourier sampling}}^{f}\text{ outputs } (\rho,1) :
\rho \in H^\perp]>\delta/|H| \ge \delta/k$.
Using these inequalities we can upper bound the
acceptance probability of the test, which is
\begin{eqnarray*}
&&\prob[\exists H \unlhd G, |H| \le k,\text{$H$ is $t$-generated} \quad
\forall i\ (b_{i}=1\implies \rho_{i}\not\in H^{\perp})]\\
&=&\prob[\exists u_1, \ldots u_t \in G,\  
\mbox{\rm Normal-closure}(\group{u_1, \ldots u_t}) = H, |H| \le k \quad
\forall i\ (b_{i}=1\implies \rho_{i}\not\in H^{\perp})]\\
&\leq& \size{G}^{t} 
\left(\max_{H\unlhd G, |H| \le k,\text{$H$ is $t$-generated}}\left\{
\prob[\mbox{\textbf{Fourier sampling}}^{f}\text{ outputs } (\rho,b) :
(b=1\implies \rho\not\in H^{\perp})]\right\}\right)^{N}\\
&<&\size{G}^{t} (1-\delta/k)^{N}
\leq 1/3.
\end{eqnarray*}
\end{proof}

\section{A classical lower bound}
Let $G$ be any Abelian group with exponent $k$.
In this section, we study the property $\range(k,1)$ for 
$k = (\log \size{G})^{O(1)}$.
We already know from Theorem~\ref{theorem:cosetrange}
that this problem has a query efficient quantum tester.
We now
prove an exponential lower bound on 
the classical testing query complexity of this problem.
Recall that the {\em exponent} of a group $G$
is the smallest integer $m$ such that $x^m = 1$ for 
every element $x \in G$. 
We prove our lower bound by adapting the proof of Theorem~4.2 of
Buhrman et al.~\cite{bfnr02}. We use Yao's minimax
principle. We construct two probability distributions $D'_1$
and $D'_2$ on the set of pairs of functions $(f_0, f_1)$,
$f_0, f_1: G \rightarrow S$,
where 
$S$ is a finite set of size $|S| = |G|^3$. Let $D'_1$ be the uniform
distribution on pairs of injective functions $(f_0, f_1)$ such
that $f_0(x) = f_1(x + u)$ for some 
element $u \in G$ and all $x \in G$. Thus, $f_0$ and $f_1$ are
$\group{u}$-similar, and $|\group{u}| \le k$.
Let $D'_2$ be the uniform distribution on
pairs of injective functions $(f_0, f_1)$ such that the ranges
$f_0(G)$ and
$f_1(G)$ are disjoint. Thus, $D'_1$ is supported on positive
instances of 
$\range(k,1)$, and $D'_2$ is supported
on negative instances of 
$\range(k,1)$ which are $1/2$-distant
from positive instances.

As in \cite{bfnr02}, instead of working with $D'_1, D'_2$, we shall
work with distributions $D_1$ and $D_2$
on pairs of functions $(f_0, f_1)$, approximating distributions
$D'_1$ and $D'_2$ respectively.
$D_1$ is got by choosing $f_1: G \rightarrow S$ and $u \in G$
independently and uniformly at random, and setting
$f_0: G \rightarrow S$ to be
$f_0(x) = f_1(x+u)$ for all $x \in G$. Since the probability that
$f_1$ is not injective is at most ${|G| \choose 2}/|G|^3 = O(1/|G|)$,
we get that $\totvar{D_1 - D'_1} = O(1/|G|)$, where
$\totvar{\cdot}$ denotes the $L_{1}$-norm or
the total variation distance. $D_2$ is got by choosing
$f_0: G \rightarrow S$ and $f_1: G \rightarrow S$ independently and
uniformly at random. The probability that at least one of $f_0, f_1$
is not injective is $O(1/|G|)$. The probability that their
ranges $f_0(G)$ and $f_1(G)$ are not disjoint is also $O(1/|G|)$.
Thus, $\totvar{D_2 - D'_2} = O(1/|G|)$.

By applying the proof technique of Theorem~4.2 of \cite{bfnr02} for
distributions $D_1, D_2$, we get the following theorem.
\begin{theorem}\label{thm:lowerbound}
Let $G$ be a finite  Abelian group and let $k$ be the exponent of $G$.
For testing
$\range(k,1)$ on $G$,
any classical randomized bounded error query algorithm 
on  $G$ requires $\Omega(\sqrt{|G|})$
queries.
\end{theorem}

\section*{Acknowledgments}
We would like to thank
Mark Ettinger, Peter H\o{}yer and G\'abor Ivanyos
for sharing their knowledge and ideas about the subject with us.

{\small

}
\end{document}